# The Near-Infrared Structure and Spectra of the Bipolar Nebulae M 2–9 and AFGL 2688: The Role of UV-pumping and Shocks in Molecular Hydrogen Excitation

Joseph L. Hora

Institute for Astronomy, 2680 Woodlawn Drive, Honolulu, HI 96822

William B. Latter

National Radio Astronomy Observatory[1], Campus Building 65, 949 N. Cherry Avenue, Tucson, AZ 85721

## ABSTRACT

High-resolution near-infrared images and moderate resolution spectra were obtained of the bipolar nebulae M 2–9 and AFGL 2688. The ability to spatially and spectrally resolve the various components of the nebulae has proved to be important in determining their physical structure and characteristics. In M 2–9, the lobes are found to have a double-shell structure. The inner shell is dominated by emission from hydrogen recombination lines, and the outer shell is primarily emission from $H_2$ lines in the 2 – 2.5 $\mu$m region. Analysis of $H_2$ line ratios indicates that the $H_2$ emission is radiatively excited. A well-resolved photodissociation region is observed in the lobes. The spectrum of the central source is dominated by H recombination lines and a strong continuum rising towards longer wavelengths consistent with a $T = 795$ K blackbody. Also present are lines of He I and [Fe II]. In contrast, the N knot and E lobe of M 2–9 show little continuum emission. The N knot spectrum consists of lines of [Fe II] and hydrogen recombination lines. In AFGL 2688, the emission from the bright lobes is mainly continuum reflected from the central star. Several molecular features from $C_2$ and CN are present. In the extreme end of the N lobe and in the E equatorial region, the emission is dominated by lines of $H_2$ in the 2–2.5 $\mu$m region. The observed $H_2$ line ratios indicate that the emission is collisionally excited, with an excitation temperature $T_{ex} \approx 1600 \pm 100$ K.

*Subject headings:* Planetary Nebulae: Individual: M 2–9 – Planetary Nebulae: Individual: AFGL 2688 – Planetary Nebulae: General – Infrared: General – ISM: Molecules – Molecular Processes

---

## 1. Introduction

Planetary Nebulae (PN) exhibit a wide range of morphologies, most of them differing from the case of a uniform spherical expanding shell. The class displaying bipolar symmetry is particularly common, including approximately 50% of all PNe (Zuckerman & Aller 1986). Most proto-planetary nebulae (PPN) have been found to be bipolar as well (see Kwok 1993). Therefore, understanding bipolar PN is an important step in learning about PN formation and evolution.

The object M 2–9 (PN G010.8+18.0) is a highly symmetric bipolar nebula, with lobes extending from opposite sides of a bright central core nearly in the plane of the sky. Bright knots of emission are visible in the lobes and at their N and S ends. The nebula is typically classed as a proto-planetary or young PN (Walsh 1981), although there are several unusual characteristics of this object that question this identification. For example, variations in the brightness, size, and position of knots of emission in both lobes have been reported in timescales of a few years (Allen & Swings 1972, van den Bergh 1974, Kohoutek & Surdej 1980). The optical spectrum of the core is also atypical of PN; Balick (1989) has noted that it has features common to objects such as $\eta$ Car, the nova RR Tel, and other stars with high mass loss. The *JHK* morphology of M 2–9 was shown by Aspin et al. (1988) to be similar to the optical structure. The knots are brighter at the shorter wavelengths, while the core is brightest at *K*. In an analysis of optical spectopolarimetry, Schmidt & Cohen (1981) find the lobes are composed of a two-phase medium; a low-ionization region filling most of the lobes, and a smaller region of higher ionization.

The object AFGL 2688 (the "Egg Nebula") is also a symmetric bipolar nebula, aligned roughly N-S with its N lobe slightly inclined towards the observer. The central source is not directly visible; the reflected optical emission from the lobes is consistent with that of a F5 supergiant (Crampton et al. 1975, Ney et al. 1975). It is an exceptionally strong source in the infrared, and has been classed as a PPN. Previous near-IR spectroscopy and imaging has shown the presence of strong $H_2$ emission from several locations in the source (Thronson 1982; Beckwith et al. 1984, Gatley et al. 1988, Smith et al. 1990, Latter et al. 1993). One remarkable structural feature of this object is an apparent "torus" of $H_2$ emission in the equatorial plane of the bipolar nebula. The orientation of the torus coincides with extended $NH_3$ (Nguyen-Q-Rieu et al. 1986) and HCN (Bieging & Nguyen-Q-Rieu 1988).

In this paper we examine the infrared emission from these two bipolar PN, M 2–9 and AFGL 2688, to determine the nature of the emission and to learn about the detailed structure of these post-AGB objects. These objects appear to represent extremes of PPN evolution, yet there are many morphological and spectral similarities between them.

## 2. Observations

### 2.1. Imaging

Near-IR images of M 2–9 were obtained with the University of Hawaii (UH) NICMOS3 camera (Hodapp et al. 1992) at the 2.2 m telescope using the f/31 secondary. Table 1 describes the filters used and images that were obtained. The images presented here were constructed of many shorter exposures which were individually sky-subtracted and flat-fielded before shifting and averaging to form the final image. The FWHM of a point source in the final images, which included all seeing and instrumental effects, was $0\rlap{.}''75$ for the *JHK* images, and $0\rlap{.}''9$ for the narrow-band images. For flux calibration, the standard stars HD 106965 and HD 162208 were used for the *J, H,* and *K* images, and HD 136754, HD 161903, and HD 203856 were used for the narrow-band images (Elias et al. 1982). The images of M 2–9 are presented in Figure 1. In the *J*-band image, the small peak at roughly (2,-2) adjacent to the main source is an artifact caused by the overexposed central source. Images of AFGL 2688 were presented in a previous paper (Latter et al. 1993).

### 2.2. Spectroscopy

The spectroscopic observations were performed during 1993 July 4-5 on the UH 2.2m telescope using KSPEC, a new near-IR spectrograph (Hodapp et al. 1994). The KSPEC instrument covers the atmospheric windows from $1 - 2.5$ $\mu$m in one spectral frame on a 256×256 pixel NICMOS3 HgCdTe detector array. The spectrograph was configured with a $0.8 \times 6.5$ arcsec slit, providing a resolution of $\lambda/\Delta\lambda \sim 700$. Images were taken simultaneously with the spectral integrations using the slit-viewing detector, allowing accurate slit placement and guiding during the integration. The slit positions for both nebulae are shown in Figure 2. The data were flat-fielded using dome flats, and the wavelength calibration was performed using spectra of an argon reference lamp. The spectra were extracted using the IRAF routines in the noao.twodspec.apextract package. Correction for atmospheric absorption was done by dividing by the spectrum of a G0 star with similar airmass (SAO 141548 for M 2–9, SAO 70834 for AFGL 2688) and then multiplying by a 5920K blackbody spectrum. The calibrated spectra are shown in Figures 3 and 4. Tables 2 and 3 list the ratios of lines detected in M 2–9 and AFGL 2688. The wavelengths and probable line identifications are given in the first two columns. The wavelength values were obtained from several sources, including Allen et al. (1985), Hamann et al. (1994), Black & van Dishoeck (1987), and Wiese et al. (1966).

The spectra were flux-calibrated using observations of standard stars. Absolute calibration is difficult because not all of the flux from the star enters the narrow slit during a single integration, and the amount differs for each integration depending on how well the star is centered on the slit (the seeing during the KSPEC observations was approximately 1″). The amount of lost light was estimated by the following method: the FWHM of the standard star was measured along the spatial direction on the slit. It was then assumed that the star is represented by a 2-D gaussian distribution with the measured FWHM, and the amount of radiation falling outside the slit was calculated for these dimensions, which was typically 20-30% of the total light. The calibration for each star was corrected by this factor, along with corrections for the airmass at the time of observation. Comparing results from different standard stars observed throughout the night indicated that this method is accurate to approximately 20%. For the observations of M 2–9 and AFGL 2688, the flux was integrated over the slit length with no correction for the slit width. There is some overlap between the $H$ and $J$ orders on the array, so for regions of extended emission (the lobe and N knot positions in M 2–9 and the E equatorial region in AFGL 2688) the overlap regions were masked in software to the following lengths: 2″ ($J$), 3″.5 ($H$), and 4″.0 ($K$). The spectra were then normalized to the length of the $K$ slit. For the core position of M 2–9 and the N lobe positions of AFGL 2688, the flux was integrated along the length of the slit to the size of the object, so no masking between orders was done and no correction was necessary.

## 3. Results

### 3.1. M 2–9 components

The regions in M 2–9 that were sampled using KSPEC were found to be spectrally distinct. The brightest component is the core, which accounts for most of the IR flux from the object, as well as providing the radiation which is scattered by the inner walls of the lobes, and excites the $H_2$ emission outside of the lobes. The measured line intensities in the various regions of the nebula are presented in Table 2.

#### 3.1.1. The core region

The brightness contrast between the core and the bipolar lobes is so large that it is difficult to image both simultaneously in a single integration time and exposure. An integration time long enough to show the faint lobes above the sky and detector noise usually saturates the detector at the position of the core. This is the case in Figure 1, where the central region is overexposed and diffraction spikes are visible from the central source in some of the images. The effect is most evident in the $K$ image, where the ratio of core to lobe intensity is the largest. To properly image the region near the core, a series of 0.1 – 0.5 second exposures were taken. These short exposures essentially "freeze" the seeing motion and result in nearly diffraction-limited images, approximately 0″.3 at 2.2 $\mu$m at the UH 2.2 m telescope, using a pixel scale of 0″.12. Figure 5 shows a profile through the core of M 2–9 and a standard star in the E-W direction, both of which were imaged at $K$ with an integration time of 0.1 seconds. The bright core is seen to be very compact at this spatial scale. The inner 0″.5 of both profiles roughly correspond, showing that the core is unresolved. However, the profile of M 2–9 is wider in the lower intensity wings, extending out to roughly 5″ from either side of the center. Comparison with the images in Figure 1 shows that this is the size of the narrow E–W waist of the nebula. The emission from the central region is therefore consistent with a point source superimposed on a fainter extended structure, approximately the same size as the waist of the bipolar nebula. There is no evidence of the reddened disk structure reported by Aspin et al. (1988), or of the 3″.4 FWHM nebula seen in the optical (e.g., Aspin & McLean 1984), although a more compact nebula may be present.

The spectrum of the core was taken with the slit aligned E-W across the source; the flux was summed over the slit so the spectrum contains flux from both the central point source and the nebula. The dominant features are the strong continuum emission increasing towards longer wavelength, and the emission lines of the hydrogen Brackett series, Pa$\beta$, [Fe II], He I, and O I. The continuum in the 1.9 – 2.5 $\mu$m range is well-fitted by a $T = 795$ K blackbody, in general agreement with previous studies (Cohen & Barlow 1974; Phillips et al. 1985); however, we detect no strong feature from the 7–10 transition of He II near 2.19 $\mu$m, and no H$_2$ is detected from the core region. The detection of H$_2$ by Phillips et al. might be due to their beam including emission from the lobes (see discussion of lobes below). The presence of strong O I emission at 1.3165 $\mu$m has been seen in several other young PN, e.g., NGC 6572 (Rudy et al. 1991a), BD+30°3639 (Rudy et al. 1991b), AFGL 618 (Latter et al. 1992; Kelly et al. 1992), and HB 12 (Rudy et al. 1993). This has been interpreted as an indication that the O I line is being excited by UV pumping from the stellar continuum.

The Brackett line ratios are in good agreement with recombination theory, as presented by Brocklehurst (1971) and Giles (1977) (Case B). Following Zhang & Kwok (1992), we use the near-IR extinction law of Draine (1990) to correct for differential extinction within the series. Assuming the line ratios as given by theory are correct, we can derive the extinction to the nebula. The following parameters were assumed: $n_e = 10^6, T = 10^4$ K (see Balick 1989). The flux of the H 4 - 11 line at 1.681 $\mu$m was not used in the fit since it coincides with a line of [Fe II]. The results of the fit are shown in Figure 6, where the corrected line ratios are plotted along with the theoretical values. The value of $E(B-V) = 0.75$ is derived (or $A_V = 2.3$ using a standard interstellar extinction law; Rieke & Lebofsky 1985). This differs slightly from the value of $E(B-V) = 0.89$ derived by Barker (1978) obtained using observations of optical lines. However, Barker's line measurements were centered on the N and S lobes excluding the core, whereas our value was derived from the core only line ratios. It is reasonable to expect the extinction to differ between these regions of the nebula.

### 3.1.2. The Lobes

The broadband *JHK* images in Figure 1 show that the lobes have a complex structure. The lobes are brightest near the middle of their extent, and fainter both towards their extreme ends and towards the core region. The brightest region forms a band across the lobe, which peaks at the outer E-W edges due to limb-brightening. There are bright knots at the extreme N and S ends of the nebula (discussed in the next section), along with two "spots" N and S of the core along the E edge of the lobe. There are additional spots and thin, filamentary emission along the edges of the lobes. Much of this structure is similar to that observed in optical images (e.g., Allen & Swings 1972, Aspin & McLean 1984).

The images presented in Figure 1 show much more detail of the nebular structure than previous optical and near-IR images because of their higher spatial resolution. The lobes of the nebula have what appears to be an inner and outer surface, with the inner side brightest at *J* and the outer edge relatively brighter at *K*. The *H*–band image clearly shows both edges. The narrowband images at 2.12 and 2.16 $\mu$m indicate the sources of emission in the *K*–band. Emission in the nebula from the $H_2$ line at 2.12 $\mu$m is seen to be almost exclusively at the outer edge, whereas the Br$\gamma$ emission at 2.16 $\mu$m is predominately from the inner edge of the nebular shell, and the knots of emission at the N and S tips of the lobes.

The relationships between the emission components are shown clearly in the color images displayed in Figure 7. In Figure 7a, the *J*, *H*, and *K* images are plotted in the colors blue, green, and red, respectively. The central source and the field star are overexposed so they appear white. The bluest components are the knots that are on the N–S axis at the tips of the lobes. The emission from the knots in the *J* image is dominated by lines of [Fe II] and Pa$\beta$. The inner parts of the lobes are blue-green, and the outer regions are red. The image in Figure 7b shows the source of the red outer edges of the lobes. In this plot the $H_2$ emission is plotted in red and the Br$\gamma$ image in green. The $H_2$ emission is seen to be primarily from the region just outside of the ionized H zone as indicated by the peak Brγ emission.

The KSPEC spectra confirm this spatial separation of emission features. As shown in Figure 3, there are differences between the core and the lobes, as well as spectral variations within the lobe itself. The lobes show a weak continuum, which is presumably because of scattering of light emitted from the core region (see §4). The spectra are dominated by two groups of lines which follow different trends in the lobes. The first group contains the recombination lines of H and He, and those of [Fe II]. This group is brightest along the inner edge of the lobe and decrease towards the outer edge. The second group of lines are those due to $H_2$ in the $K$–band region. The $H_2$ lines are fainter on the inner edge and increase towards the outer edge.

### 3.1.3. The N Knot

M 2–9 has knots placed symmetrically at the N and S poles of the lobes. The N knot is brighter and more spatially extended than the S knot. These knots are at the open ends of the lobes, presumably in direct line with the stellar wind that can travel through the low density lobe interior region relatively unimpeded. The morphology of the knot as seen in Figure 1 is suggestive of a bowshock. The brightest point of the knot is on the southern edge, centered on the N-S axis of symmetry, and the edges curve and trail upwards (N) on either side of the knot. The nebula extends well beyond the extent shown in the near-IR images — a deep optical image by Kohoutek & Surdej (1980) shows faint loops and knots at a distance of ≈1 arcmin located N and S of the nebula. The N knot exhibits H recombination line emission, but the brightest lines are those of [Fe II] and He I. The high [Fe II]/Brγ ratio (= 4.3) indicates the [Fe II] emission is due to shock excitation (e.g., Graham et al. 1987). This interpretation is consistent with the high velocities and fine-structure lines observed in the stellar spectrum by Balick (1989), and the high velocities observed by Icke, Preston, & Balick (1989). Balick also noted the similarities of the optical spectrum to that of η Car. The similarity extends to the near-IR, as evidenced by our spectra and the spectrum of η Car presented by Allen et al. (1985), and more recently by Hamann et al. (1994).

The occurrence of bright knots or "ansae" along the major axis of elliptical and bipolar PN is relatively common (see Balick 1987). The knots are characterized predominantly by emission from low-ionization lines. There are often multiple knots along the major axis, suggesting several episodes of knot formation. Balick et al. (1993) referred to these structures as fast, low-ionization emission regions (or FLIERs) because of their velocities relative to the expansion of the rest of the PN. The dynamical ages of these objects are relatively low, indicating that they were formed during recent periods of high mass loss and rapid stellar evolution.

There have been several proposed mechanisms to explain the formation of the ansae. Balick, Preston, & Icke (1987) suggested a model for elliptical PN in which interacting slow and fast winds are expanding into the remnant of an older asymmetric red giant envelope (RGE). The flow is focused along the symmetry axis leading to the formation of a higher density knot. Icke et al. (1989) extended this result to bipolar PN, although the models did not seem to have produced polar knots. Soker (1990) performed numerical simulations of the interacting winds flow and found that the focusing effect does not occur from the winds alone, but a jet from the central star could form ansae. Icke, Balick, & Frank (1992) also find that their hydrodynamic numerical simulations do not form ansae, but that the startup of the fast wind causes transient effects that may lead to their formation. They suggest that the ansae form during the RGE ejection phase when the gas is still largely molecular.

The observations of AFGL 2688 (Latter et al. 1993) and M 2–9 presented here seem to support the argument for early formation. In AFGL 2688, polar knots are observed to have formed in this young object, and are primarily $H_2$ emission (see §3.2 below). In M 2–9, little $H_2$ emission is detected; the line emission is predominantly from H and [Fe II]. The morphology and spectrum are consistent with the knot having formed earlier and being driven along by the fast stellar wind, rather than forming *in situ* at the tip of the nebular shell. The additional pair of ansae observed by Kohoutek and Surdej (1980) roughly 1 arcmin from the core along the major axis suggests that more than one ejection phase has occurred.

### 3.2. AFGL 2688

Images of the PPN AFGL 2688 at near-IR wavelengths were presented previously (Latter et al. 1993, see also Figure 2 of this paper). In all of the broad-band images (*IJHK*), the main double-lobed morphology was evident, along with two sets of "spikes" extending from the N and S ends of the lobes. The *K*–band image differs from the other broadband images in that it shows additional fainter emission to the E and W of the central position. Narrow-band continuum-subtracted images at 2.12 $\mu$m showed that this emission was from $H_2$. The 2.12 $\mu$m image also showed that there was significant $H_2$ emission at the N and S ends of the two main lobes.

The spectra presented here in Figure 4 were taken at several different positions, shown in Figure 2. Spectra were taken on the N lobe at 1″.0, 3″.0, 4″.5, 7″.0, and 8″.5 N of the center of the PPN. The spectra at positions 7″.0 and 8″.5 include radiation from the region of $H_2$ emission at the tip of the N lobe. The final position was centered on the E "lobe," in the "torus" region of the equatorial $H_2$ emission region.

### 3.2.1. AFGL 2688 Lobe Emission

All positions of the N lobe show a significant continuum emission component. The shape of the continuum varies continuously along the length of the lobe. Near the center, the intensity increases towards longer wavelengths. At the 4″.5 point, the spectrum is almost flat. At the extreme N end of the lobe, the intensity is highest at the shorter wavelengths. This is consistent with the imaging data presented by Latter et al. (1993), where profiles along the major axis of the nebula showed that the $K$–band emission from the main N and S lobes peaked closer to the nebula center, followed by the $H$ and $J$ band emission peaks further from the center. This can be understood as being a result of the relatively higher scattering at the shorter wavelengths at the extreme ends of the lobes.

The spectrum taken with the slit 3″.0 and 4″.5 N is dominated by emission from $C_2$ and CN. The $C_2$ emission is predominantly $\Delta v = 0$ ($\lambda = 1.210$ and $1.775$ $\mu$m) and $\Delta v = 1$ ($\lambda = 1.452$ $\mu$m). The other strong features are CN $\Delta v = 1$ transitions. Weaker features can be identified with CN $\Delta v = 2$ transitions. Indeed, much of the "noise" in this spectrum is due to weak molecular emission. H I absorption lines are also found, and the features in these spectra are consistent with the $H$–band spectra reported by Hrivnak, Kwok, & Geballe (1994). Similar features can be found in the spectra taken at the nearby slit positions, although much weaker. Some other emission features, such as C I might also be present, but confusion is large and identification difficult. Strong molecular carbon Swan band emission in the visible spectrum of AFGL 2688 has been known for some time (Humphreys et al. 1976; Cohen & Kuhi 1977), and is consistent with it being a very carbon-rich object. CN is also found in abundance in carbon stars, again consistent with the ancestry of AFGL 2688. Of interest, however, is the very spatially restricted area over which this emission is found, implying significantly different excitation or attenuation of the molecular features in the lobe region.

By contrast the E equatorial region shows no continuum emission, only emission lines of $H_2$ are present. This confirms the interpretation of the morphology that the lobe is located near the equatorial plane and is shielded from radiation from the central star. The line ratios observed (see §5.2 below) rule out fluorescent excitation and are consistent with shock excitation.

## 4. Models of single photon scattering in M 2–9

Polarization measurements of M 2–9 indicate that a significant amount of the continuum flux from the lobes is scattered light from the central star (Aspin & McLean 1984; Scarrott et al. 1993). Therefore, we have attempted to derive the general morphology of the dusty envelope of M 2–9 by calculating models which assume isotropic single photon scattering. This technique has been applied successfully to several other bipolar reflection nebulae (e.g. Morris 1981, Yuseh-Zadeh et al. 1984, Latter et al. 1992, Latter et al. 1993). Detailed descriptions of the modeling technique can be found in Morris (1981) and Latter et al. (1992, 1993). While we did not anticipate learning a fundamentally new result, we deemed the exercise worthwhile to facilitate direct comparisons of the distribution of scattering dust with that of other objects, and with other techniques to derive the matter distribution which have been applied to M 2–9.

M 2–9 is a complex object, with the presence of strong line emission and apparent departures from axial symmetry. For this reason, we have not attempted to derive a detailed description of the envelope morphology. However, the general form of the envelope has been reproduced. Its departure from our simple model lends insight into the origin of the unusual properties seen in M 2–9.

### 4.1. Model and assumptions

As in the previous work, no attempt has been made to match the emitted flux because of many poorly determined parameters. The relative flux between bands was considered, however. The grain albedo is fixed at the value of unity, and has no explicit wavelength dependence. Since for isotropic scattering the albedo just becomes a scaling factor, this value for the albedo does not affect our modeling results. The calculations include an inner dust free region of radius $r_0$. Such a region is inferred to be present in other objects of this type, and is expected to form as a star evolves through the post-AGB phase. Observations of thermal emission indicate that for M 2–9 $r_0 \lesssim 0\farcs 1$ (Purton et al. 1975). Changing the angular size of $r_0$ in the calculations changes the scale of the model image proportionally. Using $r_0 = 0\farcs 1$, we varied the idealized distribution of scattering dust until a satisfactory fit was found for our $J$, $H$, and $K$ band images. Comparisons were also made with data available in the literature (e.g., Scarrott et al. 1993). The models were convolved with a gaussian of FWHM = $0\farcs 75$ to simulate seeing and instrumental effects. The results are shown in Figure 8.

The general form of the scattering dust distribution which appears to best reproduce the data is given by the functional form (following Yusef-Zadeh et al. 1984)

$$f(\theta) = (1 - \sin\theta + \frac{1}{2}\sin^2\theta - \frac{1}{6}\sin^3\theta)^{\frac{1}{4}} C(\theta), \qquad (1)$$

where

$$C(\theta) = \left[1 + \left(\frac{\cos\theta_1 - \cos\theta_2}{\cos\theta - \cos\theta_2}\right)^2\right]^{-1} \quad \text{if } |\theta| < \theta_2, \qquad (2)$$

where $C(\theta) = 0$ if $\theta \geq \theta_2$, and $\theta$ is the stellar latitude such that $\theta = 0°$ at the equator. A constant outflow velocity giving a radial $r^{-2}$ density decrease is also assumed.

### 4.2. Modeling results

Fitting the model to the data gives $\theta_1 = 55°$ and $\theta_2 = 89°$. A graphical depiction of the dust distribution is plotted in Figure 9. Also plotted for comparison is the distribution of scattering grains derived for AFGL 2688 (Latter et al. 1993). Radically different functional forms were tried with minimal success. There are important differences between the model images and the data. However, the overall dust distribution of the model well describes the gross morphology of the envelope. Although the functional form of the distribution is similar to that found for the other bipolar nebulae AFGL 2688 and AFGL 618 (Latter et al. 1992; Latter et al. 1993), differences in the free parameters gives a different distribution of scattering dust.

The derived envelope morphology for M 2–9 is such that the model images are fairly insensitive to inclination angle and optical depth variations. The linear optical depth to scattering at the equator is $\tau \approx 12 \pm 3$ at $K$ and increasing to $\tau \approx 18 \pm 3$ at $J$. Because of inhomogeneities in the envelope and departures from symmetry, this value should be considered approximate. The relative brightness and circularity of the central peak indicates an inclination of the bipolar axes to the plane of the sky of $i \approx 20 \pm 10°$, with the southern lobe tilted toward the Earth. This inclination is in good agreement with other studies (Icke et al. 1989; Balick 1989).

### 4.3. Limitations of models

A number of factors have limited the success of our modeling efforts. The lack of sensitivity to scattering optical depth and inclination angle come from the particular morphology of the envelope and the relatively small size of $r_0$. For small inclinations a large value of $r_0$ creates the appearance of two bright peaks, and for a small value the peaks merge into one. If $r_0$ had been larger than $0\rlap{.}''1$ in this object, then the lack of sensitivity to $i$ would be different (see, for example, Latter et al. 1993). Multiple scattering effects in the equatorial region likely cause distortions in the observed image which cannot be reproduced by our models. More important are differences which result from apparent departures from axial symmetry, as well as a lack of perfect reflection symmetry in the North–South direction (see also Scarrott et al. 1993). We do not see evidence for departures from a constant mass loss rate over the time required to deposit the observed envelope. While there are difficulties in understanding how a single object can form a bipolar-type nebula, it is even harder to understand when the envelope departs from symmetry and homogeneity. If M 2–9 is a common envelope binary, then it is easy to envision how relatively rapid changes in the nebular emission can arise (see §6). While M 2–9 might be a binary, this does not require, however, that the formation mechanism of the bipolar nebula be a close binary system (e.g., Morris 1981).

In a study of shock propagation through model circumstellar envelopes with a variety of density distributions, Icke et al. (1989) determined a gas distribution for M 2–9 which, within the uncertainties, is in very good agreement with that found by our photon scattering models. This is especially significant because the two modeling schemes are based on very different physical processes. We note that there is no need for an over-dense disk, or "torus" around the equator of the star, which is distinct from the red giant envelope (e.g., Fig. 5 of Scarrott et al. 1993). What is required to produce the observed morphology is that the density of scattering particles decrease near the poles (see Figure 9 and Icke et al. 1989), not a density "enhancement" at the equator. It is unlikely that inclusion of multiple scattering or a detailed model of grain optical properties will change this result. The origin of non-spherically symmetric density distributions around post-AGB objects is still a mystery, so such a distinction might be important to the solution.

## 5. Excitation of Molecular Hydrogen in M 2–9 and AFGL 2688

A longstanding problem in the interpretation of H$_2$ emission has been how the near-IR transitions are excited in a given environment; collisional excitation in hot gas, shocks, or a vibrational cascade following excitation of electronic states through absorption of ultraviolet photons in the Lyman and Werner bands (fluorescence emission). Such a distinction can be very important in determining the gas kinematics of the region and the role of shocks in the shaping of PNe.

Collisionally excited emission is typically well described by a Boltzmann distribution in the ground electronic state vibration–rotation levels of the molecule. The level populations can differ significantly from a Boltzmann distribution in the presence of a strong UV field. Pure fluorescence emission from H$_2$ has been described in detail by Black & van Dishoeck (1987). The only clear case of this type of emission from a PN detected to date is Hubble 12 (Dinerstein et al. 1988). Sternberg & Dalgarno (1989) demonstrated that the characteristic fluorescence spectrum is driven to one which is very similar to that produced by collisional excitation if densities are relatively high ($n \gtrsim 10^5$ cm$^{-3}$) and the UV field is strong ($\gtrsim 10^3$ times the average interstellar flux). That work showed that even though the observed spectrum appears by the usual criteria to be shock excited, shocks produced by mass motions or winds need not be present to excite the emission. In such cases sensitive observations are required to detect the signature of fluorescence emission in lines from highly excited vibrational levels.

The wavelength dependence of attenuation caused by material in an object like a PN is likely to be much different than that found for the general interstellar medium. However, the degree to which differential extinction affects the observed spectrum can be estimated by comparing emission lines which arise from the same upper state. In the wavelength range of our data, emission lines originating in $v = 1$, $J = 2$ are the 1,0 S(0) and 1,0 Q(2) lines, and from $v = 1$, $J = 3$ are the 1,0 S(1) and 1,0 Q(3) lines. Line ratios for these transitions should simply be the product of the transition energy with the transition probability. Departures are most likely a result of differential attenuation in the spectrum, which can be corrected for in the analysis.

For a Boltzmann distribution of states, the column densities of different vibration–rotation levels are related to the $v = 1$, $J = 3$ state (origin of the 1,0 S(1) line) by

$$\frac{g(3)N(v,J)}{g(v)N(1,3)} = \exp\left(-\frac{E(v,J) - E(1,3)}{kT_{ex}}\right), \tag{3}$$

where $g(v)$ is the statistical weight. In terms of observed flux, the left side can be written as

$$= \frac{F(v',J')\nu_{1,0S(1)}A_{1,0S(1)}g(3)}{F(1,3)\nu_{\Delta v,\Delta J}A_{v',J'\to v'',J''}g(v)}, \tag{4}$$

where $A_{v',J'\to v'',J''}$ is the transition probability, and $F(v',J')$ is the observed line flux at frequency $\nu_{\Delta v,\Delta J}$. In the following, we use these relationships to derive physical parameters in the $H_2$ emitting regions.

### 5.1. $H_2$ emission in M 2–9

By comparison of emission lines from the same upper state as described above, we find that a minimum of $A_V \sim 5^{+7}_{-5}$ mag for M 2–9 is required to explain the line ratios. There is a large uncertainty due to the relatively short wavelength difference between the emission lines. In the following analysis, we have dereddened the spectra by $A_V = 5$ mag. Since the differential extinction is small over the wavelength range covered, errors introduced by departures from a pure interstellar extinction law will be small.

In Figure 10 we have plotted the natural logarithm of the observed relative populations in the O lobe slit position versus $T_u = E(v',J')/k$. If the vibration–rotation level populations were given by a strict Boltzmann distribution, all points in this diagram could be fit by a single straight line with slope equal to the negative inverse of the excitation temperature. In M 2–9 the spectra show a departure from thermodynamic equilibrium in the vibration–rotation levels. The sets of $v' \to v''$ transitions describe different slopes than if all the transitions are taken together. This indicates a different excitation temperature for the rotational levels from that of the vibrational levels. In M 2–9 the $H_2$ rotational levels have $T_{ex}(J) \approx 850 \pm 50$ K and $T_{ex}(v) \approx 2250 \pm 100$ K. This type of discrepancy in $T_{ex}$ is a clear signature of UV pumped excitation of electronic states followed by a vibrational cascade. The excitation temperature of the rotational levels should closely resemble the kinetic temperature of the gas through collisional coupling. However, the vibrational levels cannot be excited to $T_{ex}(v) \gtrsim 1000$ K by collisions alone without the interactions exciting the rotational levels to a similar temperature. The same trend is found at the other two slit positions, especially in the $v = 3 \to 2$ transitions. However, the number of detected lines is smaller, making the discrepancy in $T_{ex}$ less distinct.

It is important to note that in this case a frequently used diagnostic of fluorescent emission, the 2,1 S(1)/1,0 S(1) line ratio is $\approx 0.1$, consistent with shock excitation. However, Sternberg & Dalgarno (1989) note that this value is in the range of what they call "collisional" fluorescent emission, indicative of high density and a strong UV field. The central star of M 2–9 is classified as late O (Swings & Andrillat 1979) or early B (Calvet & Cohen 1978) and is a strong UV source. At the distance of the $H_2$ emitting region from the central star, the UV flux is expected to be $\sim 3 - 20 \times 10^5$ times that of the ambient interstellar medium flux. However, the total density of the emitting $H_2$ gas could not have been known *a priori*. Based on the results of Sternberg & Dalgarno (1989), the value is $n \gtrsim 10^5$ cm$^{-3}$. Without high sensitivity spectra, the signature of fluorescent excitation can be missed, implying the presence of shocks where none is required.

## 5.2. $H_2$ emission in AFGL 2688

The possible origins of the molecular hydrogen emission in AFGL 2688 have been discussed by Jura & Kroto (1990) and Latter et al. (1993). The data presented here are significantly more detailed and sensitive than in the previous works. The same analysis has been applied to the molecular hydrogen spectrum of AFGL 2688 as was done for for the M 2–9 data above. For these data, a visual extinction to the $H_2$ emitting region of $A_V \approx 5^{+10}_{-5}$ mag is also found. As a comparison, Thronson (1982) found $A_V \approx 3^{+11}_{-5}$ mag using only one transition pair. The dereddened data is shown in a $g(3)N(v,J)/g(J)N(1,3)$ versus $T_u$ plot in Figure 11 for the east lobe data only. There are no apparent departures from a fully collisionally-excited spectrum. A linear regression to all data points gives an excitation temperature of $T_{ex} \approx 1600 \pm 100$ K. In Figure 12, we show the data from the two north lobe positions and the east lobe plotted together. The different data sets are indistinguishable, and excitation of the molecule at all locations observed appears to be identical. From their narrowband images Latter et al. found an average column density in the north lobe of $N(H_2) \approx 2.8 \times 10^{18}$ cm$^{-2}$ assuming $T_{ex} = 2000$ K. The measured 1,0 S(1) flux is consistent with the previous work, given the difference in aperture size. Adjusting the column density from Latter et al. for the more accurate $T_{ex} = 1600$ K, we find $N(H_2) \approx 3.9 \times 10^{18}$ cm$^{-2}$ in the north lobe. The total mass of emitting $H_2$ is then $M(H_2) \approx 5.2 \times 10^{-5}$ M$_\odot$ (at $D = 1$ kpc) in the lobe region (Due to a typographical error, we note that Latter et al. should have derived $M(H_2) \approx 3.7 \times 10^{-5}$ M$_\odot$, not $M(H_2) \approx 7.6 \times 10^{-5}$ M$_\odot$.)

## 6. Internal Motions in M 2–9

The appearance of the lobes and position of the knots of emission have been reported to be variable on timescales of a few years (Allen & Swings 1972, van den Bergh 1974, Kohoutek & Surdej 1980). Kohoutek & Surdej suggest that the changes in spot position showed a rotation around the major axis of the nebula. However, it is difficult to explain how such a rotation could be established and maintained in the nebula. It would also require high velocities ($\approx 10^3$ km s$^{-1}$) which are not observed, or that the nebula is much closer than other estimates (50 pc, compared to $\gtrsim 1$ kpc). Others, such as Allen & Swings, have suggested that the walls of the nebula are being illuminated non-uniformly by the central source. A hole in the nebula surrounding the star could cause a bright spot on the walls of the lobes. This model, however, does not adequately explain the symmetry between the N and S nebular lobes.

Recently, Scarrott et al. (1993) have compared their H$\alpha$ images taken in 1992 to previous imaging of M 2–9 since 1977 and concluded that the knots have not moved since that time. They point out that the knots N2 and S2 (using the notation of Kohoutek & Surdej) are most likely the limb-brightened walls of the lobe cavity, and a variable illumination could cause a change in the position of the brightest part of the nebula. Our observations in general confirm this interpretation. The "spots" N2 and S2 are seen to be thin and elongated in the N-S direction, along the outer edge of the lobes. Corresponding limb-brighted edges are seen on the W side of the nebula as well, although not as bright. The location of the spots are wavelength-dependent; the centroid of the narrowband 2.12 and 2.16 $\mu$m images differ by several arcsec. It is likely that the positions measured in the $J$–band image also differ slightly from previous optical images, which were taken primarily in H I. In the $J$ image (Figure 1), it is evident that the edge extends into the core of the nebula, and that the spots N1 and S1 lie along this edge. These spots are also slightly extended N-S along the edge, although they are much more compact than N2 and S2. The knots N3 and S3 appear to be separate from these structures, and are not at the edge of the nebula.

The positions of the knots in our $J$–band image are given in Table 4. These are compared to the positions in previous observations in Figure 13, where the RA-Dec. positions are plotted using different symbols overlayed on a contour plot of the $J$–band image from Figure 1. In addition to the features identified in Kohoutek & Surdej (1980), the position of three additional spots, S4, S5, and S6 are indicated. These spots are most evident in the $J$ image, and barely detected in the $H$ and $K$ images. The spots are unresolved, and have no matching features in the northern lobe. The spots S4 and S5 are not associated with a lobe edge or other enhancement of emission. These may be additional regions of enhanced ionized emission. Based on the density of field stars in the region, it is also possible that these are stars superimposed on the nebula.

The significant change in the location of the spots took place between 1952 and 1971. During this period, the brightest spots apparently shifted from the W to the E side of the nebula. There is also some evidence for further slight shifts from 1971 to 1977, but since that time, the spots seem to have held their position. The best agreement is for the compact spots N1 and S1, whereas the elongated features appear less well determined. This can be understood in terms of the different seeing conditions and perhaps small changes along the length of the lobe causing the observed emission peak to shift. The wavelength observed also may effect the position of the component. For example, in the $J$ images, the peaks of emission in the N2 and S2 features were near the ends closest to the core, rather than midway along the feature length. The spots N1 and S1 are barely visible in the $K$ image. The nebula is much more symmetric in the $K$ and $H_2$ images, and the morphology is dominated by the limb-brightened edges of the lobes, rather than bright spots.

We conclude that there is little evidence for regular rotation of bright spots. Our images confirm that the present spot positions are similar to those reported in 1977. The position of the spots are resolution and wavelength dependent, which may have led to the appearance of spot rotation. The changing appearance of the nebula is consistent with irregular illumination of existing symmetric features in the lobes.

## 7. Comparison of M 2–9 and AFGL 2688

The main differences between M 2–9 and AFGL 2688 are a result of the properties of their central stars, which is related to the evolutionary state of the PN. The central stars of PN evolve from AGB stars with temperatures of $3 - 4 \times 10^3$ to more than $10^5$, as the nebula grow from a few AU to close to a parsec in size over its lifetime of $\approx 25,000$ years (e.g., see review by Kaler 1985). There are many other factors which will affect the characteristics of a PN, such as the mass of the star and nebula, the possible binarity of the star, the composition of the star and ejected shell, the stellar wind speed and its anisotropy and velocity variations over time, and the interstellar medium in which the nebula is expanding. The distribution of scattering dust as found in §4 is similar for the two objects, but differs in detail (Figure 9).

AFGL 2688 is apparently the younger of the two bipolar objects presented here, having evolved off the AGB about 200 years ago (Jura & Kroto 1990). Like M 2–9, it has exhibited relatively rapid changes, increasing in optical brightness about two magnitudes since 1920, although recently the nebula has not changed significantly (Gottlieb & Liller 1976). The central star is an F-type supergiant, with an effective temperature of $T_{eff} \approx 8000$ K (Crampton et al. 1975, Ney et al. 1975). The star temperature is much lower than the minimum needed to ionize the nebula (25,000 K, Kaler 1985). The emission from the nebula is composed primarily of reflected stellar continuum and molecular features of $C_2$, $C_3$, CN, and $H_2$ in the optical and near-IR.

M 2–9 is a more evolved object, yet at its estimated age of 5000 years it is still considered a young PN (Schmidt & Cohen 1981). Its central star is estimated to be late O or early B type, with an effective temperature of $\approx$ 30,000K (Swings & Andrillat 1979; Schmidt & Cohen 1981). The flux from the central star is therefore sufficient to ionize the nebula, and the nebular spectrum is dominated in the optical by lines ranging in ionization from O I, Fe II, [Fe II], and Si II through [O III] (Balick 1989), and in the near-IR by lines of H, He I, [Fe II], and fluorescent $H_2$ emission. A central object is directly visible; however, our simple scattering model indicates the optical depth to scattering at the equator is $\tau \sim 12$ at 2.2 $\mu$m, similar to AFGL 2688 (Latter et al. 1993). This implies that we are not seeing the central star but a pointlike region of concentrated photon scattering (Figure 8). However, the picture is far from clear.

One interesting similarity between the two nebulae are the presence of knots of shocked emission along the major axis. In AFGL 2688, there are clumps of collisionally-excited $H_2$ emission at the ends of the nebula along the major axis, located between the "spikes" seen in the images of Latter et al. (1993). In M 2–9, the knots exhibit strong [Fe II] emission, indicating higher temperature shocks. The difference between the emission of the knots in the nebulae is likely due to the higher wind velocity present in M 2–9. Knots or clumps have also been postulated to exist in the PPN AFGL 618 as well (Schmidt & Cohen 1981; Latter et al. 1992).

The molecular hydrogen emission from these two objects represents both fundamental forms of excitation mechanisms; shock or collisional excitation in AFGL 2688, and absorption of UV photons resulting in near-IR fluorescence emission in M 2–9. While the $H_2$ emission in AFGL 2688 appears to come from several different regions, the excitation of the molecule is very similar in each location. Indeed, the different data sets are indistinguishable in this regard. Latter et al. (1993) suggested that the lobe and equatorial emission did not necessarily have a common origin (i.e., a spatially correlated shock wave). These data might suggest this is not the case, and that the shocks which excite the $H_2$ result from a common wind. However, the physical conditions in the lobes and equator are different. Jura & Kroto (1990) have argued that the $H_2$ in the lobe regions can be excited by rapid grain streaming. This does not appear to be the case for the equatorial emission (Latter et al. 1993). Careful modeling of $H_2$ excitation in the rapidly evolving environment of AFGL 2688 is required to fully understand the origin of the molecular excitation. In the case of M 2–9, shocks created by winds or other mass motions are not required to excite the molecular hydrogen emission. More importantly, while wind generated shocks are present in that object, the $H_2$ emission does not trace them. It does trace, however, a region of molecular photodissociation since ≈10% of the UV photons absorbed will result in dissociation of the molecule (this is the principal route by which $H_2$ is photodissociated). A segregation of hydrogen recombination and UV excited molecular emission is clearly seen (Figure 7). The lobe regions of M 2–9 therefore represent a well-defined and resolved photodissociation region (PDR). An analogous PDR has also been observed in the young PN NGC 7027 (Graham et al. 1993).

## 8. Conclusion

We have presented new near-IR images of M 2–9 and spectra of M 2–9 and AFGL 2688. The images reveal much more detail of the structure of the nebula than has previously been observed. Some of the lobe "spots" in M 2–9 have been determined to be the limb-brightened edges of the lobes, and three new spots have been observed. The lobes exhibit a double-shelled structure; the narrow-band images and spectra show that the outer shell is dominated by $H_2$ line emission, whereas the inner edge of the lobe is primarily H recombination emission from gas in the lobe and continuum emission scattered from the central source. Models of M 2–9 were calculated using the isotropic single-photon scattering method of Morris (1981). These models constrain the global morphology of the dust envelope. There are discrepancies between the model and data which indicate departures from axial symmetry. The $H_2$ emission in the lobes of M 2–9 exhibits a spectrum consistent with UV excitation, with $T_{ex}(J) \approx 850 \pm 50$ K and $T_{ex}(v) \approx 2250 \pm 100$ K. A segregation of H recombination and $H_2$ emission is seen, showing that the PDR is clearly resolved in the lobe regions. In AFGL 2688, spectra taken along the N lobe also shows a segregation of emission components. In the spectra taken near the center of the nebula, the emission is dominated by a continuum that rises towards longer wavelengths, and molecular emission from CN and $C_2$. Moving outward along the lobe, the continuum emission is increasingly bluer, and at the tip of the lobe there is strong $H_2$ emission, as in the E equatorial region. The $H_2$ emission in the N lobe and E equatorial region appears very similar, consistent with collisional excitation with $T_{ex} \approx 1600 \pm 100$ K.

We thank D. M. Kelly for a careful reading of the manuscript and helpful comments. WBL thanks NRAO for support through a Jansky Fellowship. The KSPEC instrument project was supported in part by NASA Grant No. NAGW 755.

Fig. 1.— Images of M 2–9 obtained with the UH NICMOS3 camera at the UH 2.2m telescope. In all images, N is up and E to the left, and the lowest contour is at the 5 $\sigma$ noise level of the image. The contours are evenly spaced by the 5 $\sigma$ value for 20 contours in each image. The following are the 5 $\sigma$ values for each wavelength, in Jy arcsec$^{-2}$: a) $J$, $4\times10^{-5}$; $H$, $5\times10^{-5}$; $K$, $1.5\times10^{-4}$; b) 2.12 $\mu$m, $1.6\times10^{-4}$; 2.16 $\mu$m, $6.7\times10^{-5}$; 2.26 $\mu$m, $3\times10^{-5}$. c) Grayscale plots of the broad and narrow-band images of M 2–9.

Fig. 2.— a) $J$–band contour plot of M 2–9, with an overlay showing the slit positions of the KSPEC measurements. b) $K$–band contour plot of AFGL 2688 showing the KSPEC slit positions (image from Latter et al. 1993).

Fig. 3.— Spectra of M 2–9 taken at the positions shown in Figure 2. a) Spectra of all five positions in M 2–9. The Core spectrum is shown divided by 100; the other spectra have been offset by adding the amounts indicated in the figure. Lines of H$_2$ are marked by lines below where they appear in the E Lobe spectrum. b) Spectrum of the Core region of M 2–9, plotted on a logarithmic scale to show more detail. The detected lines of the hydrogen Brackett series are indicated by the bar below the spectrum; other lines of interest are labeled.

Fig. 4.— Spectra of AFGL 2688 taken at several positions, as shown in Figure 2. a) The 1″.0, 3″.0, and 4″.5 N positions. The 4″.5 N spectrum has been offset by 50 mJy. b) The 7″.0, 8″.5, and E equatorial spectra. The positions of detected lines of H$_2$ emission are indicated along the bottom of the plot.

Fig. 5.— Profiles in the E-W direction comparing the peak position in M 2–9 to the standard star HD 162208. The inner core of M 2–9 is unresolved; there is extended emission in the wings of the profile due to the lobes, and possibly a compact component within 1″ of the peak.

Fig. 6.— The corrected Brackett line ratios (triangles) are plotted as a function of upper level quantum number, along with the theoretical ratios (dashed line).

Fig. 7.— Color images of M 2–9 in several wavelengths. The central object and field star are overexposed and appear white. (a) Broad-band "true color" image with $J$, $H$, and $K$ plotted as blue, green, and red, respectively. (b) Two-color image, with the H$_2$ image plotted in red and the Br$\gamma$ image in green.

Fig. 8.— Model images of M 2–9 calculated assuming is isotropic single-scattering of photons from the central star. The distribution of scattering dust is shown in Figure 9. Spacing of the logarithmic contours is 0.4 dex on an arbitrary intensity scale. The models have been convolved with a gaussian of FWHM = 0″.75. (a) Model of $J$–band continuum emission (Figure 1a). (b) Model of $K$–band continuum emission (Figure 1a.). The lowest contour in (a) is 0.4 dex lower than (b). For comparison, the 5 $\sigma$ levels plotted in Fig. 1 for $J$ and $K$ differ by 0.6 dex.

Fig. 9.— The distribution of scattering dust used to calculate the models shown in Figure 8 (solid line). Angle $\theta$ is the stellar latitude, such that $\theta = 0°$ at the equator. The distribution derived for AFGL 2688 is also shown for comparison (dashed line; Latter et al. 1993).

Fig. 10.— The $H_2$ upper state vibration–rotation populations relative to that in the $v = 1$, $J = 3$ level, for the outer slit position on M 2–9. The 2,1 S(5) line is contaminated by H $n = 8 \rightarrow 4$. The observed data point is connected to a lower point from which the maximum estimated contribution of H $n = 8 \rightarrow 4$ has been removed. It is evident from these data that the vibrational excitation temperature exceeds the rotation excitation temperature, indicating fluorescence emission. Linear fits to the data finds $T_{ex}(v) \approx 2250 \pm 100$ K (dashed line) and $T_{ex}(J) \approx 850 \pm 50$ K (solid lines). The errorbars are the 1 $\sigma$ uncertainty in measured line fluxes combined quadratically. For the symbols without errorbars, the errors are smaller than the symbol size. These data have been dereddened for $A_V = 5$ mag.

Fig. 11.— Upper state $H_2$ vibration–rotation populations relative to that in the $v = 1$, $J = 3$ level, for the spectrum taken in the eastern equatorial region of AFGL 2688. These data can be fit by a single straight line indicative of equilibrium vibration–rotation level populations. A linear regression gives $T_{ex} \approx 1600 \pm 100$ K (dashed line). The errorbars are as in Fig. 10, and these data have been dereddened for $A_V = 5$ mag.

Fig. 12.— Same as Figure 11, but with all slit positions displayed (two on the north lobe, and one on the eastern equatorial emission). No lines arising from $v > 2$ are detected in the north lobe. There is no distinction in the $H_2$ excitation at the various locations. The dashed line is as in Figure 11.

Fig. 13.— Positions of "spots" in M 2–9, from 1952 to 1992. The data are from Kohoutek & Surdej 1980 (1952.4, 1971.4, 1973.58, 1977.32, and 1978.26), Scarrott et al. 1993 (1986.7 and 1992.3) and this work (1992.5). The core position was used for the reference point (0,0), and the position of ST8 marked by the cross was taken from Kohoutek & Surdej.

TABLE 1
Observations of M 2-9

| Date | Filter name | Half-power Wavelengths ($\mu$m) | Plate Scale ($''$/pix) |
|---|---|---|---|
| 1992 June 20 | J | $\cdots$ | 0.12 |
| 1992 June 21 | H | $\cdots$ | 0.12 |
| 1992 June 20 | K | $\cdots$ | 0.12 |
| 1993 June 27 | H$_2$ | 2.1132 - 2.1367 | 0.24 |
| 1993 June 27 | Br$\gamma$ | 2.1646 - 2.2053 | 0.24 |
| 1993 June 27 | 2.26 | 2.23 - 2.29 | 0.24 |

TABLE 2
M 2-9 LINE RATIOS

| Wavelength (Å) | Identification | Component Line Ratios[a] | | | | |
|---|---|---|---|---|---|---|
| | | Core | N Knot | Lobe I | Lobe M | Lobe O |
| 11880 | [Fe II], Fe I | 0.030 | 0.25 | ... | ... | ... |
| 11969 | He I | 0.030 | ... | ... | ... | ... |
| 12384 | Fe II | 0.037 | ... | ... | ... | ... |
| 12528 | He I | 0.056 | ... | ... | ... | ... |
| 12567 | [Fe II] | 0.17 | 3.5 | 0.62 | 0.11 | ... |
| 12785 | He I | 0.15 | ... | ... | ... | ... |
| 12818 | H I 3 - 5 (Pa$\beta$) | 4.08 | 3.13 | 0.61 | 0.21 | 0.14 |
| 12943 | [Fe II] | 0.034 | ... | ... | ... | ... |
| 12978 | He I | 0.018 | ... | ... | ... | ... |
| 13165 | O I | 0.27 | ... | ... | ... | ... |
| 13206 | [Fe II] | 0.031 | ... | ... | ... | ... |
| 13278 | Fe II, [Fe II] | 0.026 | ... | ... | ... | ... |
| 13616 | ? | 0.13 | ... | ... | ... | ... |
| 15191 | H I 4 - 20 | 0.028 | ... | ... | ... | ... |
| 15260 | H I 4 - 19 | 0.033 | ... | ... | ... | ... |
| 15342 | H I 4 - 18 | 0.064 | ... | ... | ... | ... |
| 15439 | H I 4 - 17 | 0.026 | ... | ... | ... | ... |
| 15556 | H I 4 - 16 | 0.058 | ... | ... | ... | ... |
| 15701 | H I 4 - 15 | 0.074 | ... | ... | ... | ... |
| 15756 | ?Fe II | 0.050 | ... | ... | ... | ... |
| 15880 | H I 4 - 14 | 0.12 | ... | ... | ... | ... |
| 15995 | [Fe II] | 0.024 | 0.31 | ... | ... | ... |
| 16109 | H I 4 - 13 | 0.098 | 0.13 | ... | ... | ... |
| 16407 | H I 4 - 12 | 0.14 | ... | ... | ... | ... |
| 16436 | [Fe II] | 0.092 | 4.3 | 1.26 | 0.42 | ... |
| 16806 | H I 4 - 11 | 0.45 | ... | ... | ... | ... |
| 16875 | Fe II | 0.21 | ... | ... | ... | ... |
| 17002 | He I | 0.075 | ... | ... | ... | ... |
| 17100 | ?[Fe II] | 0.016 | 0.14 | ... | ... | ... |
| 17354 | H I 4 - 10 | 0.34 | ... | ... | ... | ... |
| 17414 | Fe II | 0.37 | ... | ... | ... | ... |
| 19391 | ? | 0.0017 | ... | ... | ... | ... |
| 19443 | H$_2$ 2 - 1 S(5)[b] | ... | ... | 0.36 | 0.23 | 0.14 |
| 19444 | H I 4 - 8 | 0.71 | 0.85 | ... | ... | ... |
| 19535 | ? | ... | 0.19 | ... | ... | ... |
| 19570 | H$_2$ 1 - 0 S(3), [Fe II] | 0.21 | 0.12 | 0.79 | 0.56 | 0.69 |
| 19700 | H$_2$ 8 - 6 O(2) | ... | ... | 0.079 | ... | ... |
| 20332 | H$_2$ 1 - 0 S(2) | ... | 0.068 | 0.32 | 0.29 | 0.31 |
| 20418 | ? | 0.066 | ... | ... | ... | ... |
| 20581 | He I | 0.11 | 0.46 | 0.069 | 0.055 | ... |
| 20729 | H$_2$ 2 - 1 S(3) | ... | ... | 0.095 | 0.087 | 0.083 |
| 21213 | H$_2$ 1 - 0 S(1) | ... | 0.15 | 1.0 | 1.0 | 1.0 |
| 21536 | H$_2$ 2 - 1 S(2) | ... | ... | ... | 0.060 | 0.032 |
| 21655 | H I 4-7 (Br$\gamma$) | 1.0 | 1.0 | 0.33 | 0.17 | 0.051 |
| 22008 | H$_2$ 3 - 2 S(3) | ... | ... | 0.052 | 0.060 | 0.048 |
| 22227 | H$_2$ 1 - 0 S(0) | ... | ... | 0.29 | 0.25 | 0.26 |
| 22471 | H$_2$ 2 - 1 S(1) | ... | 0.030 | 0.11 | 0.097 | 0.11 |
| 23439 | H$_2$ 4 - 3 S(3) | ... | ... | ... | ... | 0.019 |
| 23550 | H$_2$ 2 - 1 S(0) | ... | ... | ... | ... | 0.034 |
| 23858 | H$_2$ 3 - 2 S(1) | ... | ... | ... | 0.049 | 0.082 |
| 24059 | H$_2$ 1 - 0 Q(1) | ... | 0.15 | 1.1 | 1.1 | 1.0 |
| 24128 | H$_2$ 1 - 0 Q(2) | ... | 0.10 | 0.39 | 0.34 | 0.34 |
| 24231 | H$_2$ 1 - 0 Q(3) | ... | 0.15 | 0.85 | 0.82 | 0.71 |

[a] The Lobe I, Lobe M, and Lobe O are the inner, middle, and outer edges of the N lobe, as described in the text. The line ratios for the Core and N Knot are relative to Br$\gamma$ fluxes of 109.5 and 1.1 $\times 10^{-14}$ erg cm$^{-2}$s$^{-1}$, respectively. The ratios for the Lobe I, Lobe M, and Lobe O components are relative to the H$_2$ 1 - 0 S(1) line at 21213 Å. The fluxes for these lines are 3.1, 4.8, and 5.2 $\times 10^{-14}$ erg cm$^{-2}$ s$^{-1}$ for the Lobe I, Lobe M, and Lobe O components, respectively. The 1 $\sigma$ noise level for these line measurements estimated from the continuum is $5 \times 10^{-16}$ erg cm$^{-2}$ s$^{-1}$.

[b] The flux for the H$_2$ 2 - 1 S(5) line is contaminated by the HI 4 - 8 line at 19446 Å.

TABLE 3
AFGL 2688 Line Ratios

| Wavelength (Å) | Identification | Component Line Ratios[a] | | |
|---|---|---|---|---|
| | | N 7″.0 | N 8″.5 | E equator |
| 16878 | $H_2$ 1 - 0 S(9) | ... | ... | 0.035 |
| 17143 | $H_2$ 1 - 0 S(8) | ... | ... | 0.019 |
| 17471 | $H_2$ 1 - 0 S(7) | 0.14 | 0.18 | 0.15 |
| 17876 | $H_2$ 1 - 0 S(6) | ... | 0.10 | 0.10 |
| 19443 | $H_2$ 2 - 1 S(5) | ... | 0.11 | 0.06 |
| 19570 | $H_2$ 1 - 0 S(3) | 1.1 | 1.2 | 1.1 |
| 20035 | $H_2$ 2 - 1 S(4) | ... | ... | 0.019 |
| 20332 | $H_2$ 1 - 0 S(2) | 0.31 | 0.33 | 0.32 |
| 20729 | $H_2$ 2 - 1 S(3) | 0.14 | 0.14 | 0.10 |
| 21213 | $H_2$ 1 - 0 S(1) | 1.0 | 1.0 | 1.0 |
| 21536 | $H_2$ 2 - 1 S(2) | ... | ... | 0.032 |
| 22008 | $H_2$ 3 - 2 S(3) | ... | ... | 0.017 |
| 22227 | $H_2$ 1 - 0 S(0) | 0.20 | 0.22 | 0.22 |
| 22471 | $H_2$ 2 - 1 S(1) | 0.091 | 0.097 | 0.083 |
| 23550 | $H_2$ 2 - 1 S(0) | ... | ... | 0.026 |
| 23858 | $H_2$ 3 - 2 S(1) | ... | ... | 0.028 |
| 24059 | $H_2$ 1 - 0 Q(1) | 0.83 | 0.79 | 0.81 |
| 24128 | $H_2$ 1 - 0 Q(2) | 0.30 | 0.33 | 0.30 |
| 24231 | $H_2$ 1 - 0 Q(3) | 0.78 | 0.73 | 0.83 |

[a]The line ratios are normalized to the $H_2$ 1 - 0 S(1) line in each lobe. The flux of this line in each of the lobes are 5.4, 7.3, and 12.0 $\times 10^{-14}$ erg cm$^{-2}$ s$^{-1}$ for the N 7″.0, N 8″.5, and E equatorial region, respectively. The 1 $\sigma$ noise level for these line measurements estimated from the continuum is $5 \times 10^{-16}$ erg cm$^{-2}$ s$^{-1}$.

TABLE 4
M 2–9: Component Positions

| Component | Offset R.A. (arcsec) | Offset Dec. (arcsec) |
|---|---|---|
| Core | 0 | 0 |
| N1 | -2.15 | 4.11 |
| N2 | -3.78 | 8.58 |
| N3 | 0.18 | 13.53 |
| S1 | -2.26 | -3.84 |
| S2 | -4.43 | -9.33 |
| S3 | -0.40 | -12.91 |
| S4 | -4.01 | -4.05 |
| S5 | 1.11 | -4.64 |
| S6 | 4.63 | -6.72 |
| ST8 | 7.90 | 3.17 |